\documentclass[nopreprintline,12pt,nonatbib]{elsarticle}

\usepackage[T1]{fontenc}
\usepackage[utf8]{inputenc}
\usepackage{capt-of}

\makeatletter
\let\c@author\relax
\makeatother

\usepackage[
  backend=biber,
  style=numeric-comp,
  sorting=none,
  url=true
]{biblatex}

\usepackage[margin=2.5cm]{geometry}

\usepackage[most]{tcolorbox}
\usepackage{xcolor}
\usepackage{placeins}
\usepackage{booktabs}

\usepackage{float}

\newcommand{\matchacommand}[2]{%
\noindent{\ttfamily #1}\par
\vspace{1.2ex}
#2\par\vspace{1.6ex}
}

\usepackage{cellspace}
\usepackage{amssymb}
\usepackage{amsthm}
\usepackage{amsmath}

\newcounter{bla}

\journal{Computer Physics Communications}

\usepackage{hyperref} 
\hypersetup{
    pdftitle={MATCHA: A Mathematica package for matching UV models onto HEFT},
    colorlinks=true,
    linkcolor=blue,
    citecolor=blue,
    urlcolor=blue,
}
\usepackage{xspace}
\newcommand{\heftmatcha}{\texttt{MATCHA}\xspace}

\usepackage{slashed}

\newcommand{\mF}{\mathcal{F}}
\newcommand{\mG}{\mathcal{G}}
\newcommand{\mL}{\mathcal{L}}

\newcommand{\be}{\begin{equation}}
\newcommand{\ee}{\end{equation}}

\newcommand{\bear}{\begin{eqnarray}}
\newcommand{\eear}{\end{eqnarray}}

\usepackage{braket}

\usepackage{mmacells}

\mmaSet{
  moredefined={
    LoadTools,
    SetSMFields,
    SetBSMFields,
    SetSMParams,
    MatchToHEFT,
    Results,
    MATCHADiagrams,
    EFTorder,
    SimplifyResult,
    ExportDiagrams,
    OutputDirectory,
    AlternativeBasis,
    OnlyRelevantDiagrams,
    MATCHA`,
    Paint,
    PaintLevel,
    Particles,
    ColumnsXRows
  }
}

\begin{document}

\begin{frontmatter}



\title{
\includegraphics[width=0.98\textwidth]{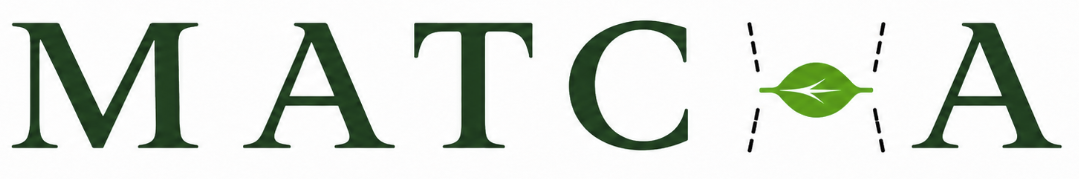}\\[1em]
MATCHA: A Mathematica package for matching UV models onto HEFT
}


\author[a]{Raquel G\'omez-Ambrosio}

\author[a]{Carlos Quezada-Calonge\corref{author}}

\cortext[author] {Corresponding author.\\\textit{E-mail address:} carlosariel.quezadacalonge@unito.it}
\address[a]{Dipartimento di Fisica, Università di Torino, and INFN, Sezione di Torino,\\ Via P.\ Giuria 1, 10125 Torino, Italy}

\begin{abstract}
We present \textbf{MATC}hing \textbf{H}EFT \textbf{A}mplitudes (\heftmatcha), a Mathematica package designed for leading-order (LO) matching of an ultraviolet (UV) model to the Higgs Effective Field Theory (HEFT). \heftmatcha performs the matching of non-decoupling effects of order $\mathcal{O}(1)$ to the LO HEFT lagrangian for an arbitrary number of Higgs fields. The main benefit of \heftmatcha is that it is built on existing packages such as \texttt{FeynArts} and \texttt{FormCalc}, which are familiar to the user, and directly benefits from the established features of these packages. In addition, \heftmatcha is designed to require minimal input from the user: solely the \texttt{FeynRules} output files and the desired order of expansion. Together, these features make \heftmatcha a practical tool for extracting the leading low-energy HEFT couplings that encode non-decoupling effects. A public release, including the code, documentation and an accompanying Mathematica notebook that reproduces the Real Singlet Extension example and serves as a template for new models, is available at \href{https://cqc-phy.github.io/MATCHA-release}{\texttt{https://cqc-phy.github.io/MATCHA-release}}.
\end{abstract}
\end{frontmatter}

\newpage


\noindent \textbf{PROGRAM SUMMARY}\\

\begin{small}
\noindent

{\em Program Title:} MATCHA                                         \\

{\em CPC Library link to program files:}
(to be added by Technical Editor)                                  \\

{\em Developer's repository link:}
\url{https://github.com/cqc-phy/MATCHA-release}                     \\

{\em Licensing provisions:}
GPLv3                                                              \\

{\em Programming language:}
Wolfram Language                                                   \\

{\em External routines/libraries used:}
Mathematica 12.0 or later, FeynArts 3.11 and FormCalc 9.8           \\

{\em Supplementary material:}
A Mathematica example notebook and the corresponding FeynArts
model files for the Real Singlet Extension.                         \\

{\em Nature of problem:}
Matching ultraviolet models onto the Higgs Effective Field Theory
and extracting the corresponding LO HEFT coefficients.                \\

{\em Solution method:}
MATCHA performs tree-level amplitude matching onto the
leading-order HEFT lagrangian. It uses FeynArts and FormCalc to
calculate the relevant ultraviolet amplitudes, expands them in the
heavy-mass limit, and extracts the non-decoupling HEFT coefficients
of order $\mathcal{O}(1)$.                                         \\

{\em Additional comments including restrictions and unusual features:}
The user must provide the FeynArts model files, identify the light
and heavy fields, and specify the heavy masses. Loop-level matching
and matching to the next-to-leading-order HEFT lagrangian are not
currently supported.

\end{small}

\newpage

\tableofcontents

\newpage

\section{Introduction}

Even with the advent of an ultimate theory that describes all existing interactions,  computational resources are finite and using such a complete theory for a specific calculation would prove impossible or highly costly. Additionally, such an approach could diminish the explainability of phenomena in terms of emergent concepts relevant to a given energy regime. Because of this, even if we were to arrive at that fundamental theory for high-energy physics, we would (probably) continue using some sort of Effective Field Theory (EFT). 

In the meantime, as we wait for that theory, 
it is important to map the effects of available UV completions to our accessible energy regime in order to ascertain whether they are theoretically possible and compatible with current experimental bounds. This procedure, of connecting the high-energy theory to a low-energy framework, is called \textit{matching}.
In this context, two effective theories exist which describe this low-energy regime: the Higgs Effective Field Theory (HEFT) and the Standard Model Effective Field Theory (SMEFT). Specifically, HEFT \cite{Longhitano:1980iz,Longhitano:1980tm,Feruglio:1992wf} is built after electroweak symmetry breaking takes place and treats the Higgs and the Goldstone bosons separately, providing the most general framework for the electroweak sector. On the other hand, SMEFT \cite{Brivio:2017vri,Isidori:2023pyp,Alonso:2013hga} is built in terms of the complex doublet of the unbroken phase of the Standard Model (SM). This embeds the Higgs and the Goldstone bosons into the same structure, the complex doublet, which has profound consequences since it generates correlated Higgs couplings \cite{Buchalla:2016bse,Gomez-Ambrosio:2022qsi,Salas-Bernardez:2022hqv,Delgado:2024mqg,Goldberg:2024eot,Domenech:2022uud,Davila:2023fkk,Anisha:2024ljc,Mahmud:2025wye,Domenech:2025nzr,Bhardwaj:2023ufl,Papaefstathiou:2023uum,Fortuna:2024rqp,Domenech:2025gmn,Lewis:2026zug}, and decoupling effects \cite{Appelquist:1974tg,Falkowski:2019tft,Buchalla:2016bse,Dawson:2023oce,Dittmaier:2026nnb,Arco:2023sac}. In this way, SMEFT is embedded into HEFT, since the latter is able to accommodate both non-decoupling and decoupling effects. 

Several software packages exist in the literature that match a given UV theory to SMEFT at various levels \cite{Fuentes-Martin:2022jrf,Carmona:2021xtq,DasBakshi:2018vni,Criado:2017khh,LopezMiras:2025gar,Guedes:2023azv,Guedes:2024vuf}, and while many models have been matched onto HEFT \cite{Buchalla:2016bse,Dawson:2023ebe,Dawson:2023oce,Arco:2023sac,Song:2025kjp,Quezada-Calonge:2026fgg}, there is no dedicated software which is able to do this in a systematic way.

\heftmatcha fills this gap by providing a remarkably user-friendly approach and requiring minimal input from the user. Schematically, the main sequence of the procedure of \heftmatcha is:

\begin{itemize}
    \item \heftmatcha already encodes the  contact amplitudes for LO HEFT, that is, the amplitudes arising from the contact terms of the processes $hh \rightarrow n \times h$, $W^+{W^-} (ZZ)\rightarrow n \times h$  and $f\bar{f} \rightarrow n \times  h$, where $W$ and $Z$ are the electroweak vector bosons, $f$ is a fermion and $h$ is the 125 GeV Higgs. 
    
    \item Then, \heftmatcha calculates the relevant UV amplitude for these processes including contact terms and propagating heavy fields from the UV lagrangian provided by the user. This amplitude is expanded in inverse powers of the heavy masses of the BSM fields and matched onto the corresponding HEFT amplitude, which provides the HEFT couplings containing the non-decoupling effects of order $\mathcal{O}(1)$.

    \item Finally, \heftmatcha exports the list of all matched HEFT couplings together with the corresponding diagrams and makes them available through the notebook interface.
\end{itemize}

\heftmatcha is able to perform the above procedure for the number of Higgs bosons in the final state defined by the "Higgs order" or Higgs multiplicity $n$ requested by the user. For example, if $n=2$ is chosen then \heftmatcha returns the coefficients up to $n=2$ for each family $\{d_3,d_4, a_1, a_2, c_1,c_2\}$ responsible for the cubic and quartic Higgs interactions, the interaction of vector bosons with one and two Higgs, and the Yukawa interactions with one and two Higgs, respectively. Because of this, \heftmatcha is naturally suited for the study of multi-Higgs processes, which have gained increasing attention in recent years as key probes of the Higgs sector and the nature of electroweak symmetry breaking, and are expected to play a central role in future colliders \cite{Gomez-Ambrosio:2022qsi,Gomez-Ambrosio:2022why,Delgado:2023ynh,Grober:2025vse,Gomez-Ambrosio:2022why,Han:2020pif,Chiesa:2021qpr}. 

This paper is structured as follows. Section~\ref{sec:heft-lag} introduces the HEFT lagrangian onto which the UV theory is matched. Sections~\ref{sec:heft-amplitudes} and~\ref{sec:uv-amplitudes} set up the HEFT and UV amplitudes entering the matching procedure implemented in \heftmatcha. Section~\ref{sec:requirements-installation} details the required packages, installation procedure, and configuration of \heftmatcha, including the linking to \texttt{FeynArts} \cite{Kublbeck:1990xc} and \texttt{FormCalc} \cite{Hahn:1998yk}. Section~\ref{sec:how-matcha-works} outlines the algorithm employed by the package. Section~\ref{sec:using-matcha} presents the functions and main features of \heftmatcha using the Real Singlet Extension (RSE) as an example. Finally, Section~\ref{sec:crosscheck} checks the results obtained for other models against those available in the literature.

\section{The HEFT lagrangian}
\label{sec:heft-lag}

In this section, we present the relevant HEFT lagrangian for the terms considered by \heftmatcha and briefly comment on several fundamental aspects of the EFT. HEFT is suitable for calculations up to a  scale $\Lambda$ typically defined as \footnote{This electroweak scale is defined in analogy with the characteristic ChPT scale $4\pi f_\pi$, where $f_\pi$ is the pion decay constant \cite{Pich:1998xt,Ecker:1988te}.}
\begin{equation}
  E \ll \Lambda \sim 4\pi v \sim 3~\mathrm{TeV}\, ,
\end{equation}
where $v$ denotes the vacuum expectation value (vev), $v=246$ GeV of the SM Higgs. The EFT is organized according to \textit{chiral dimensions}, or equivalently, powers of derivatives. This ordering allows the HEFT lagrangian to be written in the following form \cite{Dobado:1990zh,Llanes-Estrada:2017ozu,Delgado:2015kxa,Dobado:2015bir,Buchalla:2012qq,Buchalla:2015qju,Krause:2019nek, Pich:2020xzo,Pich:2012dv,Krause:2018cwe,Delgado:2017cls,Herrero:1993nc,Dobado:2019fxe}:
\begin{equation} \mathcal{L}^{\textrm{HEFT}}=\mathcal{L}_2+\mathcal{L}_4+\ldots \quad \, . 
\label{eq:heftexpansion}
\end{equation}
We refer to the chiral dimension-two term $\mathcal{L}_2$ as the leading-order (LO) HEFT lagrangian, the term of chiral dimension-four $\mathcal{L}_4$ as the next-to-leading order (NLO) HEFT lagrangian, and so on. These successive orders organize the EFT expansion in increasing powers of momentum or derivatives, providing a systematic framework to compute observables with controlled accuracy.  The power counting of the structures we need for the construction of Eq.~\eqref{eq:heftexpansion} is summarized in Table~\ref{tab:chiral-counting}.
\begin{table}[t]
\centering
\renewcommand{\arraystretch}{1.3}
\begin{tabular}{cc}
\toprule
\textbf{HEFT building block} & \textbf{Chiral counting} \\
\midrule
$W_\mu$, $Z_\mu$, $A_\mu$, $\partial_\mu$, $m_h$, $m_W$, $m_Z$, $m_t$, $m_b$, $\bar{\psi}\psi$& 1 \\
$W_{\mu\nu}$, $B_{\mu\nu}$ & 2 \\
\bottomrule
\end{tabular}
\caption{Chiral dimensions of the HEFT building blocks.}
\label{tab:chiral-counting}
\end{table}
Having established the power-counting, we can now write the bosonic part of the LO HEFT lagrangian \cite{Feruglio:1992wf,Dobado:1990zh,Llanes-Estrada:2017ozu,Delgado:2015kxa,Dobado:2015bir,Buchalla:2012qq,Buchalla:2015qju,Krause:2019nek, Pich:2020xzo,Pich:2012dv,Krause:2018cwe,Delgado:2017cls,Herrero:1993nc,Dobado:2019fxe}:
\be
\mL_{2}^{\rm bosons}
=
\dfrac{v^2}{4} \mF(h) {\rm Tr}\left\{D_\mu U^\dagger D^\mu U\right\} 
+ \dfrac{1}{2}(\partial_\mu h)^2 - V(h)   
,
\label{eq:heftdefbos}
\ee  
where $D_\mu$ is the covariant derivative
\begin{equation}
D_\mu U= \partial_\mu U + i \hat{W}_\mu U-i U \hat{B}_\mu \, . 
\end{equation}
For convenience, we also define the following gauge fields and their corresponding field strengths as
\begin{align}
\hat{B}_\mu &= g^{\prime} B_\mu \frac{\sigma^3}{2}, &
\hat{W}_\mu &= g W_\mu^a \frac{\sigma^a}{2}, \\
\hat{B}_{\mu \nu} &= \partial_\mu \hat{B}_\nu - \partial_\nu \hat{B}_\mu, &
\hat{W}_{\mu \nu} &= \partial_\mu \hat{W}_\nu - \partial_\nu \hat{W}_\mu + i[\hat{W}_\mu, \hat{W}_\nu].
\end{align}
The HEFT formalism explicitly shows the separation between the Higgs $h$ and the Goldstone bosons encoded in the matrix $U$. There exist several equivalent parametrizations of $U$ in the literature, all providing the same physical observables as they are related through a redefinition of the Goldstone fields \footnote{For a comprehensive list of different parametrizations, the reader can consult Refs.~\cite{Weinberg:1968de,Gavela:2014uta}.}. A particular case is to consider $U=1$, corresponding to the unitary gauge. This choice eliminates the Goldstone bosons and avoids the proliferation of terms arising from the non-linearity of the lagrangian, which is particularly useful in one-loop calculations. This is the limit we will adopt throughout this work. Since in HEFT the Higgs field is a singlet, the symmetry allows us to treat $\mF(h)$ and $V(h)$ in Eq.~\eqref{eq:heftdefbos} as analytical functions of the Higgs field \footnote{Note that in Eq.~\eqref{eq:flare} the notation $a=a_1/2$ is sometimes used.}:
\begin{align}
\mathcal{F}(h) & = 1 + a_1 \dfrac{h}{v} + a_2 \dfrac{h^2}{v^2} + a_3 \dfrac{h^3}{v^3}+\ldots\, ,
\label{eq:flare}
\qquad  \\
V(h) & = m_h^2 \left( \frac{h^2}{2} +\frac{d_3}{2}\dfrac{h^3}{v} +\dfrac{d_4}{8}\dfrac{h^4}{v^2} +d_5\dfrac{h^5}{v^3}+\ldots \right)\, ,
\label{eq:potential-heft}
\end{align}
where $m_h$ is the mass of the Higgs boson, and $a_1$, $a_2$, $a_3$, $d_3$, $d_4$ and $d_5$ are HEFT couplings, and the dots stand for terms with higher powers of $h$. These couplings are normalized such that the SM is recovered for $a_1=2$ and $a_2=d_3=d_4=1$, and all other couplings set to zero.

In the fermionic sector, we focus on the top and bottom quarks, which play a relevant role in the electroweak dynamics. The interaction of the Higgs with these quarks can be written as \cite{Buchalla:2020kdh,Quezada-Calonge:2022lop,Quezada-Calonge:2022rsc}
\be
\mL_{\rm 2}^{\rm fermions}
=\bar{\psi}i \slashed{D}\psi- \bar{\psi}\left(U  \mathcal{M}(h) P_R + \mathcal{M}^\dagger(h) U^\dagger P_L \right) \psi \, ,
\ee  
where $\psi=(t,b)^T$, $P_R$ and $P_L$ are the corresponding right and left chiral projectors, $\mathcal{M}(h)=\textrm{diag}(m_t,m_b) \mathcal{G}(h)$,  and the covariant derivative acting on the fermion fields is
\begin{equation}
    D_\mu \psi= \left( \partial_\mu + i g_s G_\mu + i g W_\mu P_L + i g' B_\mu (Y_L P_L+ Y_R P_R) \right) \psi \, .
\end{equation}
The function $\mathcal{G}(h)$ is an analytical function of the Higgs field,
\be
\mG(h) = 1 + c_1 \dfrac{h}{v} + c_2 \dfrac{h^2}{v^2} + \ldots \, ,
\label{eq:flare2}
\ee
with $c_1=1$ in the SM and all higher-order coefficients $c_i$ ($i \geq 2$) set to zero. In the limit $U=1$ the relevant part of this lagrangian reduces to
\be
\mL_{\rm 2}^{\rm fermions}
=
\bar{\psi}i \slashed{D}\psi - \mathcal{G}(h) \left( \, m_t \, \bar{t} t + \, m_b \, \bar{b} b \right) \, .
\ee  
Finally, the LO HEFT lagrangian relevant for our purposes is:
\begin{equation}
\mathcal{L}^{\textrm{HEFT}}=\mathcal{L}_2^{\textrm{bosons}}+\mathcal{L}_2^{\textrm{fermions}} \, .
\label{eq:HEFTlagrangian}
\end{equation}
This lagrangian provides the anomalous couplings which modify and extend the SM \footnote{In the LO basis adopted here, we do not include the custodial-symmetry-breaking operator, since it is counted at NLO.}. The first HEFT couplings are listed in Table~\ref{tab:heft-matching-coeffs}.
\begin{table}[h!]
\centering
\begin{tabular}{c|c|c}
\toprule
\textbf{Interaction} & \textbf{HEFT coupling} & \textbf{SM value} \\
\midrule
$VV h$        & $a_1$  & 2 \\
$VVhh$       & $a_2$  & 1\\
$VVhhh$       & $a_3$  & 0\\
$h^3$                  & $d_3$  & 1\\
$h^4$                  & $d_4$ & 1 \\
$h^5$                  & $d_5$ & 0 \\
$f\bar{f}h$   & $c_1$   & 1\\
$f\bar{f}hh$   & $c_2$ & 0 \\
$f\bar{f}hhh$  & $c_3$ & 0 \\
\bottomrule
\end{tabular}
\caption{HEFT interactions up to five particles and their corresponding HEFT contact terms for $V = W,Z$ and $f = t,b$.}
\label{tab:heft-matching-coeffs}
\end{table}

\section{HEFT multi-Higgs amplitudes}
\label{sec:heft-amplitudes}

In this section, we write the multi-Higgs HEFT amplitudes for the three families of couplings given by the lagrangian of Eq.~\eqref{eq:HEFTlagrangian} \footnote{It is customary to work in a basis where the fields are canonically normalized. For this reason, the HEFT couplings will be expressed in the basis of Eq.~\eqref{eq:HEFTlagrangian}. Nevertheless, after expanding the UV amplitude there is generally a term not contained in Eq.~\eqref{eq:HEFTlagrangian} involving only Higgs bosons and two powers of momenta. In order to facilitate the matching, we have considered an alternative basis containing this extra term and matched the resulting UV amplitude to it. In the end, these alternative HEFT couplings can be brought into the usual HEFT couplings of Eq.~\eqref{eq:HEFTlagrangian} through a convenient field redefinition of the Higgs field. In ~\ref{app:alternative-basis} we show explicitly how this operator enters the LO lagrangian and the required Higgs redefinition to obtain the canonical HEFT couplings.}. These families are the Higgs self-interactions $hh \rightarrow n \times h$, the gauge-Higgs interactions $W^+W^- (ZZ) \rightarrow n \times h$,  and the Yukawa interactions $f \bar{f} \rightarrow n \times h$. For $n=1$, the interactions for $h^3$, $W^+ W^- (ZZ)h$ and $q\bar{q}h$ are given by $d_3$, $a_1$ and $c_1$ and they are the same as in the UV model, since trilinear vertices are not affected by the integration of the heavy modes. For the first family, the general amplitude in HEFT arising from the contact term of Eq.~\eqref{eq:HEFTlagrangian} can be written in the following way:
\begin{equation}
\label{eq:general-Higgs-Higgs-coupling}
\mathcal{A}_{hh \rightarrow n \times h}^{\mathrm{HEFT}} =
\begin{cases}
-3\,d_3 \,\dfrac{m_h^2 }{v} \, \, , & n = 1, \\[10pt]
-3\,d_4 \,\dfrac{m_h^2 }{v^2} \, \, , & n = 2, \\[10pt]
-d_{n+2} \,(n+2)!\dfrac{m_h^2}{v^{n}} \, \,, & n \geq 3,
\end{cases}
\end{equation}
where $n$ is the number of outgoing Higgs bosons and $d_n$ is the general coupling for the Higgs self-interactions.

Similarly, the gauge-Higgs amplitudes can be written as \footnote{\heftmatcha also includes the amplitudes with two initial $Z$ vector bosons. The $W$ and $Z$ processes provide the same coupling at LO because our HEFT lagrangian does not include the custodial-symmetry-breaking operator at this order. Consequently, MATCHA can currently only match UV models that preserve custodial symmetry at LO. Here, we use the $W$ processes as an example.}:
\begin{equation}
\label{eq:general-Higgs-WW-coupling}
\mathcal{A}_{WW \rightarrow n \times h}^{\mathrm{HEFT}} =
\begin{cases}
a_1 \,\dfrac{m_W^2 }{v} \, \epsilon_1 \cdot \epsilon_2 \, , & n = 1, \\[10pt]
a_n \,n!\dfrac{m_W^2}{v^{n}} \,
 \epsilon_1 \cdot \epsilon_2 \,, & n \geq 2.
\end{cases}
\end{equation}

Finally, the Yukawa interactions are written as:
\begin{equation}
\label{eq:general-top-yukawa-coupling}
\mathcal{A}_{q\bar q \rightarrow n \times h}^{\mathrm{HEFT}} =
\begin{cases}
-\,c_1\,\dfrac{m_q}{v}\bar{v}u\,\,, & n = 1, \\[10pt]
-\,2\,c_2\,\dfrac{m_q}{v^{2}}\bar{v}u\,
\,, & n = 2, \\[10pt]
-\,n!\,c_n\,\dfrac{m_q}{v^{n}}\bar{v}u\,
\,, & n \geq 3 .
\end{cases}
\end{equation}

\section{UV multi-Higgs amplitudes}

\label{sec:uv-amplitudes}

On the UV side, \heftmatcha will calculate for each $n$ of the previous families the corresponding UV amplitude which can be decomposed in the following way:
\begin{equation}   \mathcal{A}^{\mathrm{UV}}=\mathcal{A}^{\mathrm{UV}}_{\mathrm{contact}}+\mathcal{A}^{\mathrm{UV}}_{\mathrm{light}}+\mathcal{A}^{\mathrm{UV}}_{\mathrm{heavy}}+\mathcal{A}^{\mathrm{UV}}_{\mathrm{mix}} \, ,
\end{equation}
where the label \textit{contact} refers to the contact interaction of the light modes in the UV lagrangian, \textit{light} only contains light particles as propagators, \textit{heavy} only contains heavy particles as propagators, and \textit{mix} contains both. Since we will match this amplitude to an effective lagrangian involving only local operators, we can ignore the diagrams containing \textit{light} and \textit{mix}, because they cannot be reduced to local diagrams as they involve the propagation of a light particle. Then, we only need to consider the \textit{contact} and \textit{heavy} terms, the latter involving an arbitrary number of heavy propagators and producing, after expansion, a tower of local operators. This relevant part of the amplitude can be decomposed as:
\begin{equation}
\mathcal{A}_{\mathrm{relevant}}^{\mathrm{UV}}=\mathcal{A}^{\mathrm{UV}}_{\mathrm{contact}}+\mathcal{A}^{\mathrm{UV}}_{\mathrm{heavy}} \, .
\end{equation}
In general, $\mathcal{A}_{\mathrm{relevant}}^{\mathrm{UV}}$ depends on the parameters of the UV lagrangian and the heavy masses $M_i$ of the BSM particles. As mentioned before, in order to obtain these towers of operators it is necessary to perform an expansion. The appropriate expansion for capturing the non-decoupling behavior of the EFT is the large-mass expansion. This limit, together with the choice of parameters required to avoid ill-defined couplings, has been studied in Refs.~\cite{Buchalla:2016bse,Dawson:2023ebe,Dawson:2023oce,Arco:2023sac,Buchalla:2023hqk,Brivio:2025yrr,Dittmaier:2026nnb}. Expanding this amplitude when $M_i \rightarrow \infty$ yields 
\begin{equation}
\label{eq:uv-heavy-expansion}
\tilde{\mathcal{A}}_{\mathrm{relevant}}= \tilde{\mathcal{A}}_{2}+\tilde{\mathcal{A}}_{4}+\tilde{\mathcal{A}}_{6} + \ldots \, ,
\end{equation}
where we have decomposed the expansion into a sum of terms including higher powers of chiral dimension according to Table~\ref{tab:chiral-counting}. Each term can contain contributions from $\mathcal{A}^{\mathrm{UV}}_{\mathrm{contact}}$ and $\mathcal{A}^{\mathrm{UV}}_{\mathrm{heavy}}$. Because we are working at the leading order and considering the chiral counting given by Table~\ref{tab:chiral-counting}, it is possible to show that we only need to include the terms up to two powers of momenta, since terms like $\tilde{\mathcal{A}}_{4}$ will at least be matched onto the next-to-leading order (NLO) HEFT lagrangian of chiral dimension four.

\section{Requirements and installation}
\label{sec:requirements-installation}

In order for \heftmatcha{} to calculate the amplitudes and perform the matching to HEFT, the following external packages are required. In this work, \heftmatcha{} has been tested with the package versions specified below:
\begin{itemize}
    \item \textbf{Mathematica}: version 12.0.
    \item \textbf{FeynArts} \cite{Kublbeck:1990xc}: version 3.11.
    \item \textbf{FormCalc} \cite{Hahn:1998yk}: version 9.8.
\end{itemize}

In addition, the user must provide the \texttt{FeynRules} \cite{Alloul:2013bka}
output of the UV model, namely the \texttt{.mod} and \texttt{.gen} model files
that are used as inputs by \texttt{FeynArts}. Although \heftmatcha loads the whole model and its definitions, it is recommended that the output of \texttt{FeynRules} be generated by the function \texttt{WriteFeynArtsOutput} with the option \texttt{CouplingRename $\rightarrow$ False} instead of \texttt{True}, so the dependence on the heavy masses is explicit. Additionally, as mentioned before, in order to obtain a well-defined EFT, the parameters must be chosen so they do not produce divergences. For example, three-particle vertices are part of the UV input and are not corrected by integrating out the heavy fields, so they should not diverge in the large-mass limit. For further details, see Refs.~\cite{Dawson:2023ebe,Dawson:2023oce,Brivio:2025yrr,Dittmaier:2026nnb}. \heftmatcha is available at \url{https://cqc-phy.github.io/MATCHA-release}, where further information, updates, and improvements can be found. 
The simplest way to install \heftmatcha is to clone the public repository:
\begin{tcolorbox}
\texttt{git clone https://github.com/cqc-phy/MATCHA-release}
\end{tcolorbox}
The repository also includes a Mathematica notebook with the example of the RSE, which can be used as a starting point for other UV models.

After installation, the first step is to locate the \texttt{Config.m} file and edit the association containing the paths where \texttt{FeynArts} and \texttt{FormCalc} are installed, as well as the paths to the UV model and the desired output directory for \heftmatcha:
\begin{tcolorbox}
\texttt{\$MATCHARoutesConfig = <| \\
"FeynArtsRoute"   -> "/path/to/FeynArts", \\
"FormCalcRoute"   -> "/path/to/FormCalc", \\
"ModelPath" -> "/path/to/Models", \\
"OutputDirectory"    -> "/path/to/MATCHA" \\
|>;
}
\end{tcolorbox}
Once these paths are correctly set, \heftmatcha{} is ready to use. The main commands of \heftmatcha{} are summarized in Table~\ref{tab:matcha-commands}.

\section{How \heftmatcha works}
\label{sec:how-matcha-works}

The workflow of \heftmatcha is straightforward as it calculates one pair of amplitudes at a time in the following order: self-Higgs interactions, gauge-Higgs interactions and finally Yukawa interactions. In each case, the
Higgs order $n$ denotes the Higgs multiplicity of the process, namely the
number of Higgs bosons in the final state. The input \texttt{higgsOrderMax}
sets the maximum multiplicity considered and therefore determines the HEFT
couplings computed by \heftmatcha. The calculation then proceeds as follows:

\begin{itemize}

      \item Within each family, \heftmatcha{} starts from the Higgs order $n=1$ and proceeds up to the maximum order specified by the user, \texttt{higgsOrderMax} \footnote{It is important to note that, besides the technical issues of multiparticle processes, for a certain $n$ the approximation in perturbation theory may break down and the results are no longer reliable.}. Only after completing one family does it move on to the next one.

    \item For every HEFT amplitude, \heftmatcha{} generates the corresponding UV diagrams with \texttt{FeynArts}, using the model file provided by the user. The separation between light and heavy contributions is determined from the fields specified through \texttt{SetSMFields} and \texttt{SetBSMFields}. This defines the following set of diagrams: \texttt{RelevantDiagrams}, obtained by excluding internal SM fields, and therefore including contact diagrams and diagrams containing the heavy BSM fields to be integrated out, \texttt{IgnoredDiagrams}, obtained by keeping only light propagating fields, and \texttt{AllDiagrams}, containing the complete set.

    \item The relevant diagrams are then sent to \texttt{FormCalc} \footnote{The main limitation of \heftmatcha emerges when HEFT couplings with a large number of Higgs bosons are requested, as the computational cost and internal structure of \texttt{FeynArts} and \texttt{FormCalc} become increasingly demanding.}, which computes the analytical UV amplitude associated with each process. In this way, every HEFT amplitude is paired with the corresponding UV amplitude.

    \item The UV amplitudes are expanded in the heavy-mass limit, using the BSM masses given by the user. After this expansion, the result is rewritten in terms of the same kinematic and Lorentz structures appearing in the corresponding HEFT amplitude, so that both expressions can be compared term by term.

    \item The matching equations are obtained by equating the expanded UV amplitudes to the HEFT amplitudes. These equations are solved for the HEFT coefficients appearing in that basis.

    \item Once the coefficients have been obtained, \heftmatcha{} performs the Higgs field redefinition needed to remove the non-canonical kinetic terms. This brings the result to the canonical HEFT basis.

    \item The final Mathematica output contains the HEFT coefficients in the canonical basis in a table and can be accessed individually by requesting a particular coupling, for example, \texttt{Results["a3"]}. In addition, \heftmatcha{} exports the HEFT coupling expressions both in Mathematica format and in JSON format. Finally, if the option \texttt{ExportDiagrams -> True} has been used, the corresponding Feynman diagrams are exported to the chosen output folder.

\end{itemize}

After the calculation is completed, the diagrams associated with each HEFT coupling can be accessed through \texttt{MATCHADiagrams}. For example, the relevant diagrams for \texttt{"a1"} can be displayed with
\[
\texttt{Paint[MATCHADiagrams["a1"]["RelevantDiagrams"]]}.
\]
The last entry can also be replaced by \texttt{"IgnoredDiagrams"} or \texttt{"AllDiagrams"}.

\section{Using \heftmatcha with the Real Singlet Extension (RSE)}
\label{sec:using-matcha}

In this section, we describe how to use \heftmatcha to perform the matching of a UV model onto HEFT. To demonstrate the workflow, we will consider one of the simplest extensions of the scalar sector: the Real Singlet Extension (RSE) \cite{Buchalla:2016bse,Robens:2015gla,Robens:2016xkb,Dawson:2023oce,Boggia:2016asg,Dittmaier:2026nnb}. More specifically, we focus on its $Z_2$-symmetric version. The model extends the SM Higgs doublet with a real singlet $S$, odd under a discrete $Z_2$ symmetry. The potential can be written as
\begin{equation}
V = -\frac{\mu_1^2}{2}\,\phi^\dagger \phi - \frac{\mu_2^2}{2} S^2 
+ \frac{\lambda_1}{4} (\phi^\dagger \phi)^2 + \frac{\lambda_2}{4} S^4 
+ \frac{\lambda_3}{2}\, \phi^\dagger \phi\, S^2 \, ,
\end{equation}
where all parameters are real. The fields can be parametrized in the following way
\begin{equation}
\phi = 
\begin{pmatrix}
G^+ \\
\dfrac{1}{\sqrt{2}}(v+h_1+iG^0)
\end{pmatrix}\, ,
\qquad
S = \dfrac{v_s+h_2}{\sqrt{2}} \, ,
\end{equation}
with $v=246$ GeV and $v_s$ the singlet vev. Minimizing the potential imposes the following relations on the parameters of the lagrangian
\be
\label{eq:Z2RSE_minima}
\mu_1^2=\dfrac{\lambda_1 v^2+\lambda_3 v_s^2}{2}\, ,
\qquad
\mu_2^2=\dfrac{\lambda_3 v^2+\lambda_2 v_s^2}{2} \, .
\ee
We obtain the physical fields through the matrix
\begin{equation}
\begin{pmatrix}
h \\ H
\end{pmatrix}
=
\begin{pmatrix}
\cos\chi & -\sin\chi \\
\sin\chi & \cos\chi
\end{pmatrix}
\begin{pmatrix}
h_1 \\ h_2
\end{pmatrix}\, ,
\end{equation}
where $h$ is identified with the 125 GeV Higgs boson, $H$ is a BSM heavy scalar, and $\chi$ is the mixing angle. The RSE model is defined by the following set of parameters:
\begin{equation}
\{ v,\, m,\, v_s,\, M,\, s_\chi \} \, ,
\label{eq:1-indep-set-Z2RSE}
\end{equation}
with $m$ and $M$ being the masses of the SM Higgs and the heavy $H$ boson, and $s_{\chi} \equiv \sin(\chi)$.

This model is encoded in the model file \texttt{"Singlet\_for\_MATCHA"}, provided with the package, which we will refer to as the user-defined model that includes both $s_\chi$ and $c_{\chi} \equiv \cos(\chi)$ in order to simplify the final expression. The particle content of the singlet model is shown in Fig.~\ref{fig:rse-classes}, where we observe that the SM Higgs corresponds to the scalar index "15". This choice illustrates that the user-defined index number is irrelevant since it will be provided to \heftmatcha.

The accompanying notebook, available in the repository, provides a ready-to-run version of the following workflow. \heftmatcha is loaded from a Mathematica notebook. First, move to the directory where the package is located
\begin{tcolorbox}[
  colback=gray!10,
  colframe=gray!50,
  boxrule=0.5pt,
  breakable
]
\begin{mmaCell}{Input}
SetDirectory["/path/to/MATCHA"];
\end{mmaCell}

\begin{mmaCell}{Output}
"/path/to/MATCHA"
\end{mmaCell}
\end{tcolorbox}
Then load the package. If installed correctly, the following header will be displayed
\begin{tcolorbox}[
  colback=gray!10,
  colframe=gray!50,
  boxrule=0.5pt,
  breakable
]
\begin{mmaCell}{Input}
  << MATCHA\textasciigrave{}
\end{mmaCell}
\begin{mmaCell}{Output}
  ========================================\\
  \textbf{MATCHA}\\
  MATChing H(EFT) Amplitudes\\
  by Raquel G\'omez-Ambrosio and Carlos Quezada-Calonge\\
  Version: MATCHA 1.0\\
  ========================================
\end{mmaCell}
\end{tcolorbox}
The required external packages, \texttt{FeynArts} and \texttt{FormCalc}, are loaded using the \texttt{LoadTools} command.
\begin{tcolorbox}[
  colback=gray!10,
  colframe=gray!50,
  boxrule=0.5pt,
  breakable
]
\begin{mmaCell}{Input}
  LoadTools[]
\end{mmaCell}
\begin{mmaCell}{Output}
  Loading FeynArts...

  FeynArts 3.11 (2 Sep 2019)

  by Hagen Eck, Sepp Kueblbeck, and Thomas Hahn

  Loading FormCalc...

  FormCalc 9.8 (22 Apr 2019)

  by Thomas Hahn

  Tools loaded successfully.
\end{mmaCell}
\end{tcolorbox}

At this stage, the environment is ready to perform the matching onto HEFT. In order to translate the information from the model to \heftmatcha, the first step is to define the required SM fields for the desired processes. These include the Higgs, the gauge boson $W$ (or $Z$), the charged Goldstone boson $G^+$, the neutral Goldstone boson $G^0$, and the quark field used in the Yukawa matching, which in this example is the top quark $t$, as defined in the model. It is not necessary to specify all the particles, but only those entering the processes to be matched. If both $W$ and $Z$ are provided, \heftmatcha uses the $W$ process for the gauge-Higgs matching. By inspecting the relevant particle content, we are able to specify the mentioned fields:
\FloatBarrier
\begin{tcolorbox}[
  colback=gray!10,
  colframe=gray!50,
  boxrule=0.5pt,
  breakable
]
\begin{mmaCell}{Input}
SetSMFields[<|"Higgs" -> \{"S", 15\},\\
"GaugeCharged" -> \{"V", 2\},\\
"GoldstoneCharged" -> \{"S", 5\},\\
"GoldstoneNeutral" -> \{"S", 3\},\\
"Quark" -> \{"F", 1\}|>]
\end{mmaCell}

\begin{mmaCell}{Output}
SM fields set to: <|Higgs -> S(15),\\
GaugeCharged -> V(2),\\
GoldstoneCharged -> S(5),\\
GoldstoneNeutral -> S(3),\\
Quark -> F(1)|>
\end{mmaCell}
\end{tcolorbox}
\FloatBarrier

Next, we specify the heavy BSM fields that are being integrated out. In this case, the only BSM field is the heavy scalar $H$ from Fig.~\ref{fig:rse-classes}:

\begin{tcolorbox}[
  colback=gray!10,
  colframe=gray!50,
  boxrule=0.5pt,
  breakable
]
\begin{mmaCell}{Input}
SetBSMFields[\{\{"S", 6\}\}]
\end{mmaCell}

\begin{mmaCell}{Output}
BSM fields set to: \{S(6)\}
\end{mmaCell}
\end{tcolorbox}

To help simplify the final expression \footnote{This is not strictly necessary for the scalar and gauge sectors. For Yukawa matching, \texttt{"QuarkMass"} should be set to the mass of the selected quark. Otherwise, the generic \texttt{Mq} may appear in the result.}, we can also provide the notation used in the user-defined model file for the variables used in Eqs.~\eqref{eq:general-Higgs-Higgs-coupling}, \eqref{eq:general-Higgs-WW-coupling} and \eqref{eq:general-top-yukawa-coupling}:
\FloatBarrier
\begin{tcolorbox}[
  colback=gray!10,
  colframe=gray!50,
  boxrule=0.5pt,
  breakable,
  left=2pt,right=2pt,top=2pt,bottom=2pt
]
\begin{mmaCell}{Input}
SetSMParams[<|"HiggsMass" -> m,\\
"SMvacuum" -> v,\\
"WMass" -> Mw,\\
"QuarkMass" -> Mt|>]
\end{mmaCell}

\begin{mmaCell}{Output}
SM parameters set to:\\
<|HiggsMass -> m,\\
SMvacuum -> v,\\
WMass -> Mw,\\
QuarkMass -> Mt|>
\end{mmaCell}
\end{tcolorbox}
\FloatBarrier
Finally, we run the matching through the function \texttt{MatchToHEFT} and provide the name of the model as it appears in the directory, or only the \texttt{.mod} file name when the \texttt{.gen} file has the same name, together with the maximum Higgs multiplicity $n$ and the masses associated with the heavy fields listed in \texttt{SetBSMFields}. For the case of the RSE, the mass of the heavy scalar $H$ is $M$, and we will calculate the matching up to the Higgs order $n=3$:
\FloatBarrier
\begin{tcolorbox}[
  colback=gray!10,
  colframe=gray!50,
  boxrule=0.5pt,
  breakable
]
\begin{mmaCell}{Input}
MatchToHEFT["Singlet\_for\_MATCHA", 3, \{M\}, \\
ExportDiagrams -> True,\\
OnlyRelevantDiagrams -> True]
\end{mmaCell}

\begin{mmaCell}{Output}
MATCHA: Matching "Singlet\_for\_MATCHA" model onto HEFT\\

[1/7] Generating UV diagrams ...\\
[2/7] Generating UV amplitudes ...\\
[3/7] Processing UV amplitudes ...\\
[4/7] Expanding UV amplitudes ...\\
[5/7] Rewriting kinematics ...\\
[6/7] Solving ...

[7/7] Obtaining HEFT couplings ...
\end{mmaCell}
\end{tcolorbox}

\FloatBarrier

\begin{table}[t]
\centering
\renewcommand{\arraystretch}{1.3} 
\begin{tabular}{|Sc|Sc|}
\hline
\textbf{HEFT coefficient} & \textbf{Matching Expression} \\

\hline

$d_3$ &
$\displaystyle c^3 - \frac{s^3 v}{v_s}$ \\

$d_4$ &
$\displaystyle
\frac{
3 c^6 v_s^2
-16 c^4 s^2 v_s^2
-38 c^3 s^3 v v_s
-16 c^2 s^4 v^2
+3 s^6 v^2
}{
3 v_s^2
}$ \\

$d_5$ &
$\displaystyle
\frac{
c^2 s^2 (c v_s + s v)^2
\left(
-c^3 v_s
-2 c^2 s v
+2 c s^2 v_s
+s^3 v
\right)
}{
2 v_s^3
}$ \\

$a_1$ &
$\displaystyle 2c$ \\

$a_2$ &
$\displaystyle
-\frac{
c\left(
c s^2 v_s
- c v_s
+ s^3 v
\right)
}{
v_s
}$ \\

$a_3$ &
$\displaystyle
-\frac{
c^2 s^2
\left(
c^3 v_s^2
+5 c^2 s v v_s
+c s^2(4v^2 - 3v_s^2)
+3 c v_s^2
-3s(s^2-1)v v_s
\right)
}{
3 v_s^2
}$ \\

$c_1$ &
$\displaystyle c$ \\

$c_2$ &
$\displaystyle
-\frac{
c s^2(c v_s + s v)
}{
2 v_s
}$ \\

$c_3$ &
$\displaystyle
-\frac{
c^2 s^2
\left(
c^3 v_s^2
+5 c^2 s v v_s
+c s^2(4v^2 - 3v_s^2)
-3s^3 v v_s
\right)
}{
6 v_s^2
}$ \\
\hline
\end{tabular}
\caption{HEFT couplings for the RSE with $Z_2$ symmetry obtained with \heftmatcha.}
\label{tab:coeffs_Z2RSE}
\end{table}
Once the \texttt{MatchToHEFT} routine is finished, \heftmatcha will provide a table similar to Table~\ref{tab:coeffs_Z2RSE} with the HEFT coefficients up to order $n=3$, in agreement with the results presented in Refs.~\cite{Buchalla:2016bse,Dawson:2023oce,Quezada-Calonge:2026fgg}.

It is possible to access the expressions for each HEFT coupling. For example, if we are interested in the coupling $a_2$:
\begin{tcolorbox}[
  colback=gray!10,
  colframe=gray!50,
  boxrule=0.5pt,
  breakable,
  left=2pt,right=2pt,top=2pt,bottom=2pt
]
\begin{mmaCell}{Input}
Results["a2"]
\end{mmaCell}

\begin{mmaCell}{Output}
 c^2 - c^2*s^2 - c*s^3*v/v_s 
\end{mmaCell}
\end{tcolorbox}
We observe the expressions conform to what is expected of non-decoupling HEFT as they should not depend on the heavy BSM mass $M$. These non-decoupling corrections start at order $\mathcal{O}(1)$, so they remain when the limit $M\rightarrow \infty$ is taken. Checking the results also helps us to determine if there are different written expressions for the same BSM mass, as there are often instances where the same mass is written in different forms, such as $M$ or $M2$ in the user-defined model. \heftmatcha is able to handle both the mass and the mass squared at LO, but we recommend using the expression written in terms of the mass, such as $M$, to make the heavy-mass expansion explicit and avoid ambiguities.

\begin{figure}[t]
  \centering
\includegraphics[width=0.5\textwidth,
trim=0pt 0pt 0pt 100pt,
clip]{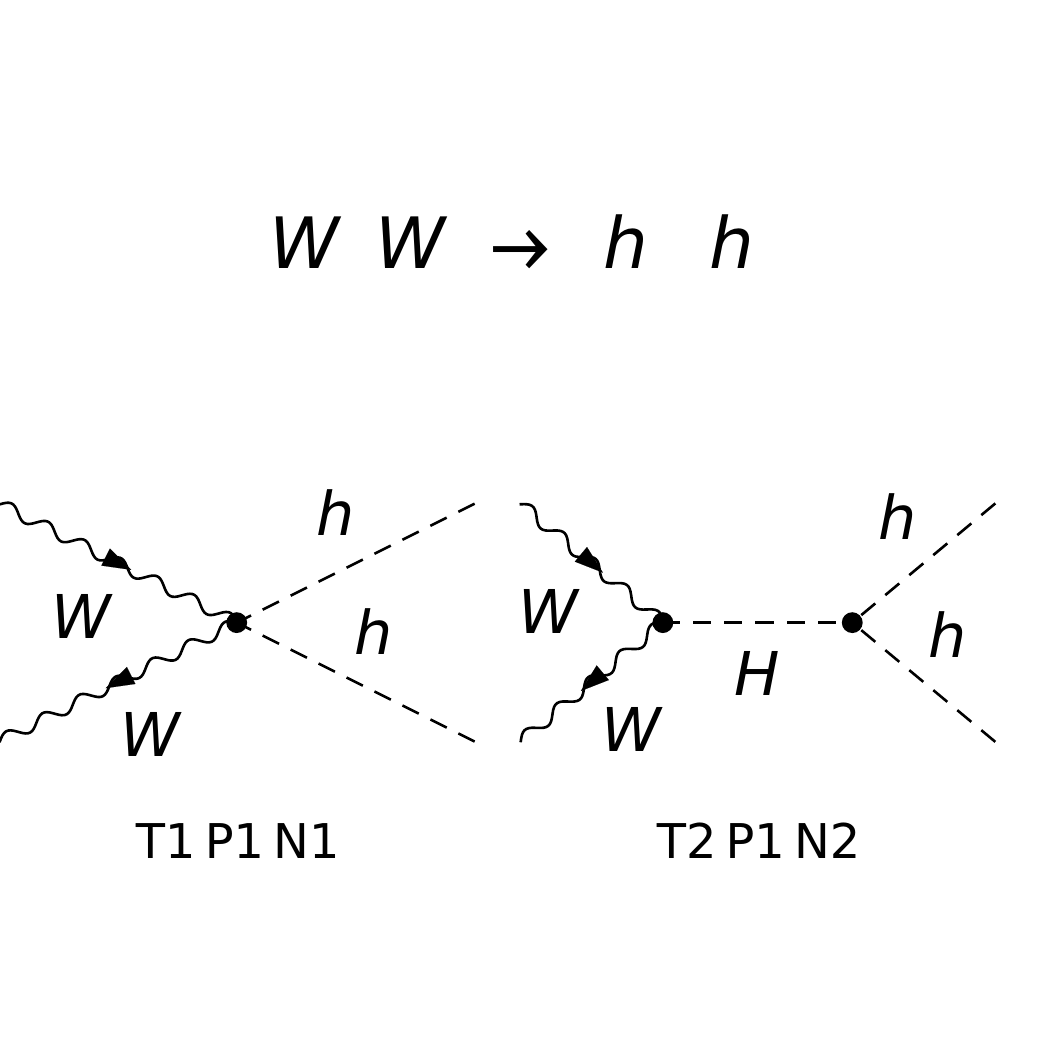}
\caption{Relevant UV diagrams for the RSE contributing to $a_2$.}
  \label{fig:a2_relevant}
\end{figure}

\begin{figure}[H]
  \centering
\includegraphics[width=0.8\textwidth]{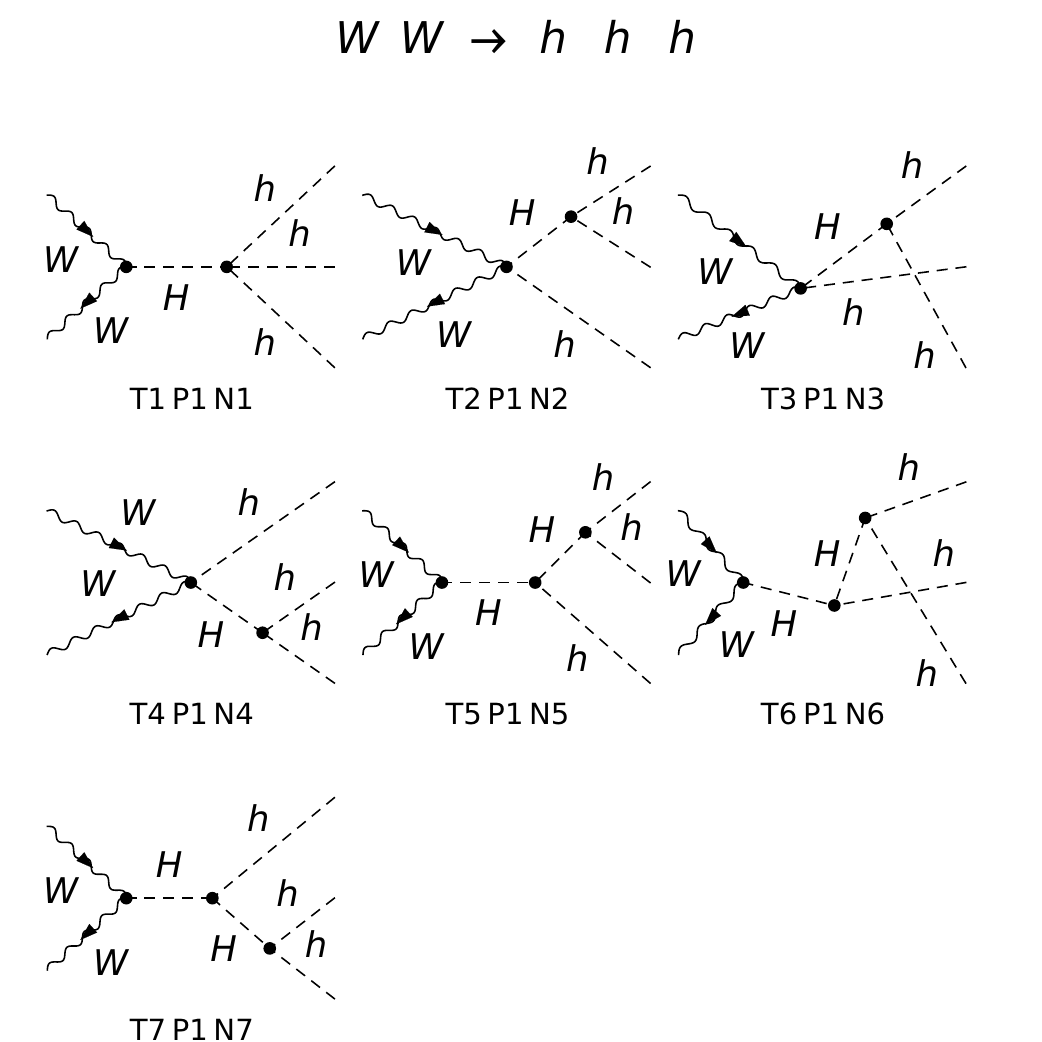}
\caption{Relevant UV diagrams for the RSE contributing to $a_3$.}
  \label{fig:a3_relevant_RSE}
\end{figure}

\begin{figure}[t]
\centering
\includegraphics[width=0.4\textwidth]{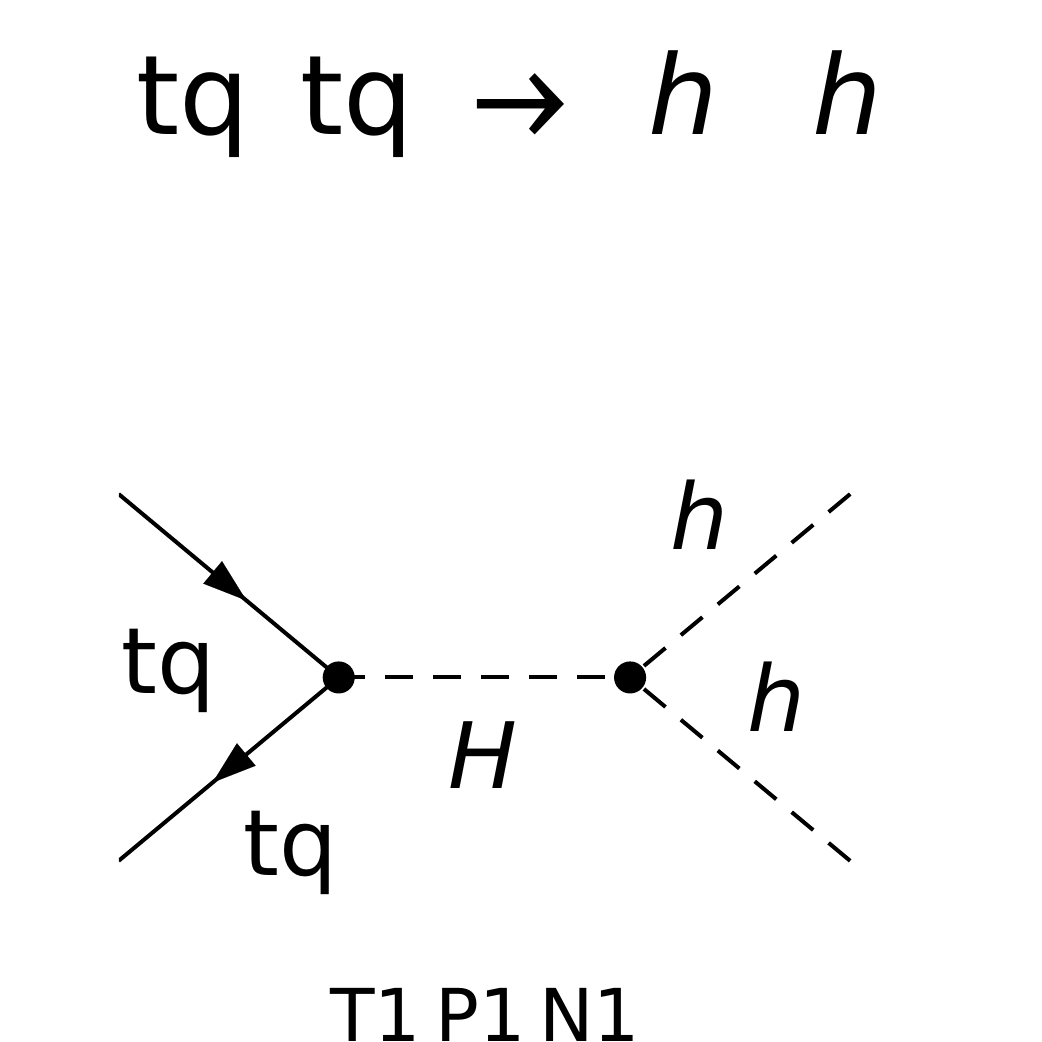}
\caption{Relevant UV diagrams for the RSE contributing to $c_2$.}
\label{fig:c2_relevant_RSE}
\end{figure}

In addition, \heftmatcha allows the user to inspect the diagrams involved in the calculation in order to assess whether all necessary SM and BSM fields have been included in the functions \texttt{SetSMFields} and \texttt{SetBSMFields}. As discussed earlier, these diagrams are divided into the categories: All, Relevant and Ignored. The Mathematica notebook enables the user to access the usual functions of \texttt{FeynArts} and use them to visualize, modify and export the diagrams:
\begin{tcolorbox}[
  colback=gray!10,
  colframe=gray!50,
  boxrule=0.5pt,
  breakable,
  left=2pt,right=2pt,top=2pt,bottom=2pt
]
\begin{mmaCell}{Input}
Options[Paint]
\end{mmaCell}

\begin{mmaCell}{Output}
\{PaintLevel -> InsertionLevel,\\
ColumnsXRows -> 3,\\
AutoEdit -> True,\\
SheetHeader -> Automatic,\\
Numbering -> Full,\\
FieldNumbers -> False,\\
DisplayFunction :> (Print /@ Render(##1) &)\}
\end{mmaCell}
\end{tcolorbox}
For example, to inspect the diagrams contributing to the HEFT coefficient $a_2$, we can use
\begin{tcolorbox}[
  colback=gray!10,
  colframe=gray!50,
  boxrule=0.5pt,
  breakable,
  left=2pt,right=2pt,top=2pt,bottom=2pt
]
\begin{mmaCell}{Input}
Paint[MATCHADiagrams["a2"]["RelevantDiagrams"],\\
PaintLevel -> \{Particles\},\\
ImageSize -> 500,\\
ColumnsXRows -> 3]
\end{mmaCell}
\end{tcolorbox}
and \heftmatcha will produce Fig.~\ref{fig:a2_relevant},
where the UV contact diagram and the heavy $H$ exchanged in the $s$-channel can be seen. If we are interested in the next gauge coupling $a_3$, the diagrams contributing to this HEFT coefficient are shown in Fig.~\ref{fig:a3_relevant_RSE}. For higher multiplicities, we would have to run \texttt{MatchToHEFT} again with at least $n=4$. \heftmatcha will produce the corresponding table, although at this point visualizing the diagrams can be cumbersome due to their number. To illustrate another family of HEFT couplings, Fig.~\ref{fig:c2_relevant_RSE} shows the relevant diagrams contributing to the Yukawa coefficient $c_2$, where the heavy scalar enters the matching of the fermionic HEFT coupling. Finally, if the matching had been run with the option \texttt{OnlyRelevantDiagrams -> False}, the remaining diagram categories can be inspected by replacing \texttt{"RelevantDiagrams"} with \texttt{"AllDiagrams"} or \texttt{"IgnoredDiagrams"}. For the coupling $a_2$, this corresponds to
\[
\texttt{MATCHADiagrams["a2"]["AllDiagrams"]}
\]
and
\[
\texttt{MATCHADiagrams["a2"]["IgnoredDiagrams"]}.
\]
For these examples, the ignored diagrams for $a_2$ and $c_2$ are shown in Figs.~\ref{fig:a2_ignored} and~\ref{fig:c2_ignored}.

\section{Cross-checks}
\label{sec:crosscheck}

Having explained how \heftmatcha works with a simple model, let us recover known results for the scalar sector of other models using \heftmatcha and the corresponding model files.

\subsection{Complex Singlet Extension (CSE)}

A straightforward extension of the RSE is the Complex Singlet Extension (CSE) \cite{Barger:2008jx,Dawson:2023oce}, in which the SM scalar sector is enlarged by a complex singlet field $S_c$. Unlike in the RSE, no discrete $Z_2$ symmetry is imposed \cite{Dawson:2017jja}, allowing for a more general scalar potential. The scalar interactions are described by the potential
\begin{eqnarray}
V &=& -\dfrac{\mu^2}{2} \phi^{\dagger} \phi+\dfrac{\lambda}{4}\left(\phi^{\dagger} \phi\right)^2 + \dfrac{1}{2} b_2  \left|S_c\right|^2 + \dfrac{\delta_2}{2}  \phi^{\dagger} \phi  \left|S_c\right|^2 + \dfrac{1}{4} \rho_2 \left(\left|S_c\right|^2\right)^2 \nonumber \\
&&+ \Bigg[ \hat{a}_1 S_c + \dfrac{1}{4} b_1 S_c^2 + \dfrac{1}{6} e_1 S_c^3 + \dfrac{1}{6} e_2 S_c\left|S_c\right|^2 + \dfrac{1}{8} \rho_1 S_c^4+\dfrac{1}{8} \rho_3 S_c^2\left|S_c\right|^2 \nonumber \\
&&+ \dfrac{1}{4} \delta_1 \phi^{\dagger} \phi S_c + \dfrac{1}{4} \delta_3 \phi^{\dagger} \phi S_c^2 + \mathrm{h.c.} \Bigg]\, ,
\end{eqnarray}
where $\mu^2$, $\lambda$, $\rho_2$, $\delta_2$, and $b_2$ are real, and the remaining couplings are in general complex. The fields can be parametrized as
\be
\phi=\left(\begin{array}{c}
G^{+} \\
\dfrac{1}{\sqrt{2}}\left(v + h + i G_0\right)
\end{array}\right)\, , \qquad
S_c= \dfrac{S + i A}{\sqrt{2}} \, ,
\ee
so that the scalar spectrum contains the usual Goldstone bosons, the scalar $h$ and the additional scalar fields $S$ and $A$. The minimization of the potential provides the relations:
\be
\mu^2 = \dfrac{\lambda}{2} v^2 \, ,
\qquad
\hat{a}_1 = - \dfrac{\delta_1}{8} v^2 \, .
\ee
After the potential is minimized, the CP-even and CP-odd components mix with $h$, giving rise to three mass eigenstates $h_1$, $h_2$ and $h_3$, where $h_1$ is identified with the observed 125 GeV Higgs boson. The mixing matrix can be written as
\be
\begin{pmatrix}
h_1 \\[4pt]
h_2 \\[4pt]
h_3
\end{pmatrix}
=
\begin{pmatrix}
c_{\theta_1} & -s_{\theta_1} & 0 \\
s_{\theta_1} c_{\theta_2} & c_{\theta_1} c_{\theta_2} & s_{\theta_2} \\
s_{\theta_1} s_{\theta_2} & c_{\theta_1} s_{\theta_2} & -c_{\theta_2}
\end{pmatrix}
\begin{pmatrix}
h \\[4pt]
S \\[4pt]
A
\end{pmatrix}\, ,
\ee
with mixing angles $\theta_1$ and $\theta_2$. Taking $\lambda, \delta_1, b_1$ and $b_2$ as dependent parameters, their expressions in terms of the independent parameters read:
\begin{subequations}
\label{eq:CSE-deprels}
\begin{eqnarray}
\label{eq:CSE-lambda}
{\lambda} &=& \dfrac{2}{v^2} \,  \Big[ m_1^2 \, c_{\theta_1}^2 + s_{\theta_1}^2 \,  \left( m_2^2 \, c_{\theta_2}^2 + m_3^2 \, s_{\theta_2}^2 \right) \Big] , \\
{\delta_{1\mathrm{R}}} &=& \dfrac{\sqrt{2} \, s_{\theta_1} c_{\theta_1}}{v} \Big[m_2^2 + m_3^2 - 2 \, m_1^2 + \left( m_2^2 - m_3^2 \right) (c_{\theta_2}^2 - s_{\theta_2}^2) \Big], \\
{\delta_{1\mathrm{I}}} &=& \dfrac{2 \sqrt{2} \, s_{\theta_1} s_{\theta_2} c_{\theta_2}}{v} \left(m_3^2 - m_2^2 \right), \\
{b_{1\mathrm{R}}} &=& - \dfrac{\delta_{3\mathrm{R}} \, v^2}{2} + m_1^2 \, s_{\theta_1}^2 + c_{\theta_2}^2 \left(m_2^2 \, c_{\theta_1}^2 - m_3^2\right) + s_{\theta_2}^2 \left(m_3^2 \, c_{\theta_1}^2 - m_2^2\right), \\
{b_{1\mathrm{I}}} &=& 2 \, c_{\theta_1} s_{\theta_2} c_{\theta_2} \left(m_3^2 -m_2^2\right) - \dfrac{\delta_{3\mathrm{I}} \, v^2}{2}  , \\
{b_2} &=& m_1^2 \, s_{\theta_1}^2 + c_{\theta_2}^2 \left( m_2^2 \, c_{\theta_1}^2 + m_3^2\right) + s_{\theta_2}^2 \left( m_3^2 \, c_{\theta_1}^2 + m_2^2\right) - \dfrac{\delta_{2} \, v^2}{2} ,
\end{eqnarray}
\end{subequations}
where we used the notation $x_{\mathrm{R}} = \mathrm{Re}(x)$ and $x_{\mathrm{I}} = \mathrm{Im}(x)$. The model is defined by the following set of parameters:
\be
\{v, m,  M, s_{\theta_1}, \theta_2,  \delta_2, \delta_3,  \rho_1, \rho_2,  \rho_3,  e_1 , e_2\} \, ,
\ee
where $m$ is the mass of the SM Higgs, $M=m_2=m_3$ corresponds to the heavy scalar mass scale in the degenerate case for the two BSM scalars, $s_{\theta_1} = \sin (\theta_1)$, and $c_{\theta_1} = \cos (\theta_1)$.

Now that we have explained the model, we begin the process of matching by translating the information about the SM fields from the CSE model file to \heftmatcha. We will omit showing our model file and just state the required dictionary:
\begin{tcolorbox}[
  colback=gray!10,
  colframe=gray!50,
  boxrule=0.5pt,
  breakable,
  left=2pt,right=2pt,top=2pt,bottom=2pt
]
\begin{mmaCell}{Input}
SetSMFields[<|\\
"Higgs" -> \{"S", 4\},\\
"GaugeCharged" -> \{"V", 2\},\\
"GoldstoneCharged" -> \{"S", 5\},\\
"GoldstoneNeutral" -> \{"S", 3\}
|>]
\end{mmaCell}

\begin{mmaCell}{Output}
SM fields set to:\\
<|Higgs -> S(4),\\
GaugeCharged -> V(2),\\
GoldstoneCharged -> S(5),\\
GoldstoneNeutral -> S(3)|>
\end{mmaCell}
\end{tcolorbox}
Next, we specify the BSM fields:
\begin{tcolorbox}[
  colback=gray!10,
  colframe=gray!50,
  boxrule=0.5pt,
  breakable,
  left=2pt,right=2pt,top=2pt,bottom=2pt
]
\begin{mmaCell}{Input}
SetBSMFields[\{\{"S", 6\}, \{"S", 7\}\}]
\end{mmaCell}

\begin{mmaCell}{Output}
BSM fields set to:\\
\{S(6), S(7)\}
\end{mmaCell}
\end{tcolorbox}
The model parameters are translated as follows:
\begin{tcolorbox}[
  colback=gray!10,
  colframe=gray!50,
  boxrule=0.5pt,
  breakable,
  left=2pt,right=2pt,top=2pt,bottom=2pt
]
\begin{mmaCell}{Input}
SetSMParams[\{"HiggsMass" -> mh,\\
"SMvacuum" -> v,\\
"WMass" -> Mw \}]
\end{mmaCell}

\begin{mmaCell}{Output}
SM parameters set to:\\
<|HiggsMass -> mh,\\
SMvacuum -> v,\\
WMass -> Mw|>
\end{mmaCell}
\end{tcolorbox}
Finally, we perform the matching by providing the name of the model, the desired order of the matching and the masses of the heavy fields.
\begin{tcolorbox}[
  colback=gray!10,
  colframe=gray!50,
  boxrule=0.5pt,
  breakable,
  left=2pt,right=2pt,top=2pt,bottom=2pt
]
\begin{mmaCell}{Input}
MatchToHEFT["CSE_MATCHA", 2, \{m2, m3\}]
\end{mmaCell}
\end{tcolorbox}

Once the calculation is finished, \heftmatcha provides the results of Table~\ref{tab:coeffs_CSE}.  As in the RSE example, the results are independent of the heavy mass scale and agree with the expressions available in the literature \cite{Dawson:2023oce,Quezada-Calonge:2026fgg}.
\begin{table}[t!]
\centering
\renewcommand{\arraystretch}{1.8}
\begin{tabular}{|Sc|Sc|}
\hline
\textbf{HEFT coupling} & \textbf{HEFT expression} \\
\hline

$d_3$ &
\parbox{0.78\textwidth}{
\centering
$\dfrac{c_{\theta_1}}{4}(c_{\theta_1}^2 - 3s_{\theta_1}^2 + 3)
-\dfrac{v}{24 m_h^2}
\bigg[3 c_{\theta_1} v \big(c_{\theta_1}^2-3 s_{\theta_1}^2-1 \big) \bar{\delta }_{23}$\\
$\qquad + \sqrt{2} s_{\theta_1} \big(-3 c_{\theta_1}^2+s_{\theta_1}^2+3 \big) e_{12 \rm{R}}\bigg]$
} \\

$d_4$ &
\parbox{0.78\textwidth}{
\centering
$c_{\theta_1}^4\left(1 - \dfrac{19s_{\theta_1}^2}{3}\right)
+\dfrac{s_{\theta_1}^2 v}{2 m_h^2}
\bigg[s_{\theta_1}^2 v \Big(-6 (c_{\theta_1}^2+1) \bar{\delta }_{23}
+\rho_{13 \rm{R}}+\rho_2 \Big)$\\
$\qquad +6 v \bar{\delta }_{23}
- 2 \sqrt{2} c_{\theta_1}^3 s_{\theta_1} e_{12 \rm{R}}\bigg]$
} \\
$ a_1$ & $2c_{\theta_1}$ \\

$a_2$ & $c_{\theta_1}^4 $ \\
\hline
\end{tabular}
\caption{HEFT couplings for the CSE obtained with \heftmatcha.}
\label{tab:coeffs_CSE}
\end{table}

\subsection{Two-Higgs Doublet Model (2HDM)}

The scalar sector of the 2HDM considered here contains two $SU(2)_L$ doublets, $\Phi_1$ and $\Phi_2$, whose vevs are defined as $v_1/\sqrt{2}$ and $v_2/\sqrt{2}$, respectively, and are taken to be real. A softly broken $Z_2$ symmetry is imposed, under which $\Phi_1 \to \Phi_1$ and $\Phi_2 \to -\Phi_2$. We introduce an angle $\beta$ such that $t_\beta = v_2/v_1$, which allows us to rotate to the Higgs basis~\cite{Donoghue:1978cj,Georgi:1978ri,Botella:1994cs,Branco:1999fs}:
\be
\label{eq:basis-rot}
\left(\begin{array}{c}
H_{1} \\
H_{2}
\end{array}\right)=\left(\begin{array}{cc}
c_{\beta} & s_{\beta} \\
-s_{\beta} & c_{\beta}
\end{array}\right)\left(\begin{array}{c}
\Phi_{1} \\
\Phi_{2}
\end{array}\right) \, ,
\ee
where only $H_1$ acquires a vev, $\langle H_1 \rangle = v / \sqrt{2}$ with $v \equiv \sqrt{v_1^2 + v_2^2} = 246$ GeV, while $H_2$ has none. In the Higgs basis, the lagrangian becomes \cite{Donoghue:1978cj,Georgi:1978ri,Botella:1994cs,Branco:1999fs}
\begin{subequations}
\begin{eqnarray}
\label{eq:kinetic}
\mathcal{L}_{\mathrm{kin}} &=& \left(D_{\mu} H_1\right)^{\dagger} \left(D^{\mu} H_1\right) + \left(D_{\mu} H_2\right)^{\dagger} \left(D^{\mu} H_2\right),\\[3mm]
\label{eq:potential}
V &=& Y_1 H_{1}^{\dagger} H_{1}
+ Y_2 H_{2}^{\dagger} H_{2}+\left(Y_3 H_{1}^{\dagger} H_{2}+\textrm{h.c.}\right) \nonumber \\
&&+ \frac{Z_{1}}{2}\left(H_{1}^{\dagger} H_{1}\right)^{2}+\frac{Z_{2}}{2}\left(H_{2}^{\dagger} H_{2}\right)^{2}+Z_{3}\left(H_{1}^{\dagger} H_{1}\right)\left(H_{2}^{\dagger} H_{2}\right)+Z_{4}\left(H_{1}^{\dagger} H_{2}\right)\left(H_{2}^{\dagger} H_{1}\right) \nonumber \\
&& + \left\{\frac{Z_{5}}{2}\left(H_{1}^{\dagger} H_{2}\right)^{2}+Z_{6}\left(H_{1}^{\dagger} H_{1}\right)\left(H_{1}^{\dagger} H_{2}\right)+Z_{7}\left(H_{2}^{\dagger} H_{2}\right)\left(H_{1}^{\dagger} H_{2}\right)+ \textrm{h.c.}\right\}.
\end{eqnarray}
\end{subequations}
The minimization of the potential imposes the relations
\be
\label{eq:theYs}
Y_1 = - \dfrac{Z_1}{2} v^2,
\qquad
Y_3 = - \dfrac{Z_6}{2} v^2 \, .
\ee
In general, the parameters $Y_3$, $Z_5$, $Z_6$, and $Z_7$ may be complex. Here, however, we focus on the CP-conserving limit, in which all of them are real. The scalar fields can then be parametrized as
\begin{align}
\label{eq:Higgs_basis_param}
H_1 = 
\begin{pmatrix}
G^+ \\
\frac{1}{\sqrt{2}}(v + h_1^{\mathrm{H}} + i G_0)
\end{pmatrix},
\hspace{3mm}
H_2 = 
\begin{pmatrix}
H^+ \\
\frac{1}{\sqrt{2}}(h_2^{\mathrm{H}} + i A)
\end{pmatrix} \, ,
\end{align}
where $h_1^{\mathrm{H}}$, $h_2^{\mathrm{H}}$, $G_0$, and $A$ are real fields, while $G^+$ and $H^+$ are complex. Except for $h_1^{\mathrm{H}}$ and $h_2^{\mathrm{H}}$, these fields already correspond to mass eigenstates: $G_0$ and $G^+$ are the Goldstone bosons, whereas $A$ and $H^+$ denote the pseudoscalar and charged scalar bosons, respectively. The mass matrix in the CP-even sector, spanned by $h_1^{\mathrm{H}}$ and $h_2^{\mathrm{H}}$, can be diagonalized by introducing a mixing angle $\alpha$ such that
\be
\label{eq:diagonalization}
\left(\begin{array}{c}
h \\
H
\end{array}\right)
=
\left(\begin{array}{cc}
s_{\beta-\alpha} & c_{\beta-\alpha}\\
c_{\beta-\alpha} & - s_{\beta-\alpha}
\end{array}\right)
\left(\begin{array}{c}
h_1^{\mathrm{H}}\\
h_2^{\mathrm{H}}
\end{array}\right) \, ,
\ee
where $h$ and $H$ are the neutral scalar mass eigenstates, with $h$ corresponding to the Higgs boson observed at the LHC and $H$ a heavy BSM neutral boson. Defining the masses of the physical states $h, H, A$ and $H^{\pm}$ to be $m_{h}, m_{H}, m_A$ and $m_{H^{\pm}}$, we can choose the following independent parameters \cite{Arco:2023sac,Buchalla:2023hqk,Dawson:2023oce} to describe the non-decoupling regime of the 2HDM, with $m_{12}$ defined as in Ref.~\cite{Arco:2023sac}:
\begin{equation}
\label{eq:indep-real-m12}
\{c_{\beta \! - \! \alpha},
\,
\beta,
\,
v,
\,
m_{h},
\,
m_{12},
\,
m_{H},
\,
m_A,
\,
m_{H^{\pm}}\}\, .
\end{equation}
We begin by translating the information about the SM fields from the 2HDM to \heftmatcha:
\begin{tcolorbox}[
  colback=gray!10,
  colframe=gray!50,
  boxrule=0.5pt,
  breakable,
  left=2pt,right=2pt,top=2pt,bottom=2pt
]
\begin{mmaCell}{Input}
SetSMFields[<|"Higgs" -> \{"S", 4\},\\
"GaugeCharged" -> \{"V", 2\},\\
"GoldstoneCharged" -> \{"S", 5\},\\
"GoldstoneNeutral" -> \{"S", 3\}|>]
\end{mmaCell}
\end{tcolorbox}

\begin{figure}[t] \centering \includegraphics[width=0.8\textwidth,
trim=0pt 100pt 0pt 0pt,
clip]{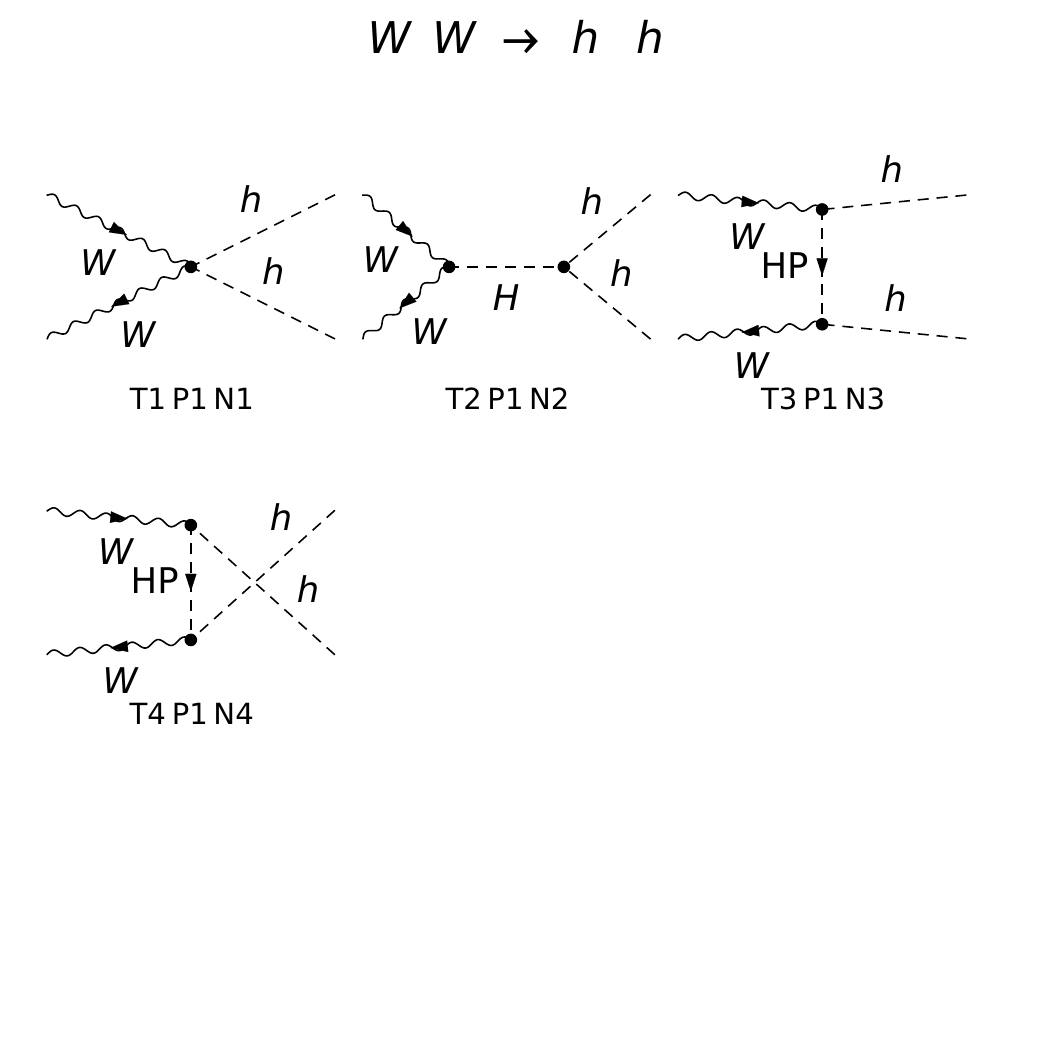} \caption{Relevant UV diagrams for the 2HDM contributing to $a_2$.}\label{fig:a2_relevant_2HDM} \end{figure}

Next, we specify the BSM fields:
\begin{tcolorbox}[
  colback=gray!10,
  colframe=gray!50,
  boxrule=0.5pt,
  breakable,
  left=2pt,right=2pt,top=2pt,bottom=2pt
]
\begin{mmaCell}{Input}
SetBSMFields[\{\{"S", 6\}, \{"S", 7\}, \{"S", 8\}\}]
\end{mmaCell}

\begin{mmaCell}{Output}
BSM fields set to:\\
\{S(6), S(7), S(8)\}
\end{mmaCell}
\end{tcolorbox}
Finally, we perform the matching by providing the name of the model, the desired order of the matching and the masses of the heavy fields.
\begin{tcolorbox}[
  colback=gray!10,
  colframe=gray!50,
  boxrule=0.5pt,
  breakable,
  left=2pt,right=2pt,top=2pt,bottom=2pt
]
\begin{mmaCell}{Input}
MatchToHEFT["2HDM_MATCHA", 2, \{mH, mHP, mA\}]
\end{mmaCell}
\end{tcolorbox}
Once the calculation is finished, \heftmatcha provides the results of Table~\ref{tab:coeffs_2HDM}, which agree with those available in the literature \cite{Arco:2023sac,Buchalla:2023hqk,Dawson:2023oce,Quezada-Calonge:2026fgg} and show the expected non-decoupling form.
\begin{table}[t!]
\centering
\renewcommand{\arraystretch}{2.0}
\begin{tabular}{|c|c|}
\hline
\textbf{HEFT coupling} & \textbf{HEFT expression} \\
\hline

$d_3$ 
&\rule{0pt}{5.5ex} $ s_{\beta-\alpha}
+ \dfrac{2 c_{\beta-\alpha}^2 c_{\beta+\alpha}}{s_{2\beta}}
\left(1-\dfrac{m_{12}^2}{m_h^2 s_{\beta} c_\beta} \right)
$ \\
$d_4$ 
& $1-\dfrac{c_{\alpha -\beta }^2}{s_{2 \beta }^2}
\left[
\dfrac{1}{6}
\left(-12 c_{2 (\alpha +\beta )}-19 c_{4 \alpha }+7\right)
-\dfrac{m_{12}^2
\left(-2 c_{2 \alpha } c_{2 \beta }-3 c_{4 \alpha }+1\right)}
{c_{\beta } m_h^2 s_{\beta }}
\right]$ \\
$ a_1$ 
& $2s_{\beta-\alpha} $ \\
$ a_2$ 
& $1+c_{\beta-\alpha}^2
\Big(1 - 2 c_{\beta-\alpha}^2
+ 2 c_{\beta-\alpha} s_{\beta-\alpha} \cot 2\beta\Big)
$ \\[6pt] 
\hline
\end{tabular}
\caption{HEFT couplings for the 2HDM obtained with \heftmatcha.}
\label{tab:coeffs_2HDM}
\end{table}
To complete the 2HDM example, we inspect the diagrams to check how the heavy charged fields enter the matching. As an example, Fig.~\ref{fig:a2_relevant_2HDM} shows the relevant diagrams contributing to $a_2$, where the charged scalars also contribute, as expected.

\FloatBarrier

\clearpage

\section{Conclusions}

In this work, we have presented \heftmatcha, a dedicated Mathematica package designed to perform the $\mathcal{O}(1)$ matching to the HEFT lagrangian at LO in the EFT for an arbitrary number of Higgs bosons. We have outlined the structure and functionality of the package in detail, describing its routines and illustrating its use through a comprehensive example based on the Real Singlet Extension. Furthermore, we have used \heftmatcha to obtain the HEFT coefficients resulting from the matching of the Complex Singlet Extension and the 2HDM, finding full agreement with the literature. These results validate the algorithm of \heftmatcha and showcase the simplicity of its implementation. In addition, we provide a Mathematica notebook that guides the user through the workflow and the commands necessary to reproduce the results and to start the matching of a new model.

This package is particularly relevant for the study of multiple Higgs processes, especially as experimental measurements become more precise and systematically probe these interactions. The ability to survey a variety of models and extract the non-decoupling effects for an arbitrary number of Higgs bosons may help assess the realization of the electroweak sector and identify the UV model lying above the electroweak scale. This software is a stepping stone towards building a HEFT package including NLO tree-level matching as well as one-loop effects.

\section*{Acknowledgements}

We are grateful to Juan José Sanz Cillero, Rafael L. Delgado, and Alexandre Salas-Bernárdez for very fruitful discussions and valuable comments. Our work has been supported by the Università di Torino under the grant DataSMEFT23 (PNRR - EU next generation) and INFN Iniziativa Specifica SPIF.

\newpage

\appendix

\section{Alternative HEFT LO basis}
\label{app:alternative-basis}

Although the HEFT basis described in Eq.~\eqref{eq:heftdefbos} is the most commonly used, as it is non-redundant, it is not the most general one since we have omitted the following operator $\mathcal{P}(h/v)$: 
\be
\mL_{2}^{\rm bosons}
=
\dfrac{v^2}{4} \mF(h) {\rm Tr}\left\{D_\mu U^\dagger D^\mu U\right\} 
+ \dfrac{1}{2}\bigg(1+2 \, \mathcal{P}\left(\frac{h}
{v}\right)\bigg)(\partial_\mu h)^2 - V(h)   
\label{eq:heftdefbosK} \, ,
\ee
with the analytical polynomial 
\be
\mathcal{P}\left(\frac{h}{v}\right)=p_2 \frac{h^2}{v^2}+p_3 \frac{h^3}{v^3}+ \ldots \, .
\ee

In this basis, the amplitude expression for the pure Higgs family is:
\begin{equation}
\label{eq:general-Higgs-Higgs-coupling-alternative}
\mathcal{A}_{hh\rightarrow n\times h}^{\mathrm{HEFT}} =
\begin{cases}
-3\,d_3\,\dfrac{m_h^2}{v} \, ,
& n = 1, \\[10pt]

\dfrac{2\,p_{2}}{v^2}\,\mathcal{K}_2
-3\,d_4\,\dfrac{m_h^2}{v^2} \, ,
& n = 2, \\[10pt]

\dfrac{n!\,p_{n}}{v^n}\,\mathcal{K}_n
-d_{n+2}\,(n+2)!\,\dfrac{m_h^2}{v^n} \, ,
& n \geq 3 ,
\end{cases}
\end{equation}

\begin{equation}
\label{eq:Kn-definition}
\mathcal{K}_n =
S_{12}
+
\sum_{a=3}^{n+1}
\left(
T_{1a}+T_{2a}
\right)
+
\sum_{3\leq a<b\leq n+1}
S_{ab}
-
(n-2)\sum_{i=1}^{n+1}k_i^2 ,
\qquad n\geq 2 .
\end{equation}
The quantity $\mathcal{K}_n$ collects the momentum dependence generated by the derivative operator. The invariants $S_{ab}$ and $T_{ia}$ denote the kinematic
combinations built from the external momenta, with $i=1,2$ labeling the two incoming
Higgs bosons and $a,b=3,\dots,n+1$ labeling the outgoing ones. Then, in order to obtain the
canonical HEFT coupling, it is possible to perform a redefinition of the Higgs field of the
form:
\be
h'=\int_0^h \sqrt{1+ 2 \, \mathcal{P}(s/v)} ds \, .
\ee
Accordingly, \heftmatcha matches onto the lagrangian of Eq.~\eqref{eq:heftdefbosK} including the term
$\mathcal{P}(h/v)$, automatically performs the redefinition to arrive at the lagrangian of
Eq.~\eqref{eq:heftdefbos} and then extracts the HEFT couplings. Nonetheless, if the user
wishes to obtain the HEFT couplings in the basis of Eq.~\eqref{eq:heftdefbosK},
\heftmatcha can also provide them.

\section{Main commands of \heftmatcha}
\label{sec:commands-list}

We summarize in Table~\ref{tab:matcha-commands} the main \heftmatcha commands used to define the model input, perform the matching and access the resulting HEFT couplings and diagrams.

\begin{center}
\begin{minipage}{0.95\linewidth}

\begin{tcolorbox}[
  colback=gray!20,
  colframe=black,
  width=0.95\linewidth,
  boxrule=0.5pt,
  arc=0pt,
  left=8pt,right=8pt,top=6pt,bottom=6pt
]

\fontsize{10.5pt}{13.0pt}\selectfont

\matchacommand
{LoadTools[]}
{Loads \texttt{FeynArts} and \texttt{FormCalc} using the routes configured in \texttt{Config.m}.}

\matchacommand
{SetSMFields[assoc]}
{Sets the SM field map. The input may be an Association or a list of rules.
The allowed keys are \texttt{"Higgs"}, \texttt{"GaugeCharged"},
\texttt{"GaugeNeutral"}, \texttt{"GoldstoneCharged"},
\texttt{"GoldstoneNeutral"} and \texttt{"Quark"}. The value assigned to each key
must be of the form \texttt{\{"S", n\}}, \texttt{\{"V", m\}},
\texttt{\{"F", k\}}, or \texttt{\{"F", k, \{gen\}\}}, specifying the
corresponding FeynArts field type and index defined by the user.}

\matchacommand
{SetBSMFields[fieldList]}
{Defines the mapping of the heavy BSM fields to be integrated out. The input
\texttt{fieldList} must be a list containing entries of the form
\texttt{\{"S", n\}}, \texttt{\{"V", m\}}, \texttt{\{"F", k\}},
\texttt{\{"F", k, \{gen\}\}}, or \texttt{\{"U", r\}}, where the first element
specifies the field type and the second one specifies the corresponding FeynArts
index defined by the user.}

\matchacommand
{SetSMParams[assoc]}
{Sets the SM parameter map. The input may be an Association or a list of rules,
and may contain any subset of the keys \texttt{"HiggsMass"},
\texttt{"SMvacuum"}, \texttt{"WMass"} and \texttt{"QuarkMass"}. For Yukawa
matching, \texttt{"QuarkMass"} should be set to the mass of the selected quark.
Otherwise, the generic \texttt{Mq} may appear in the result.}

\matchacommand
{MatchToHEFT[modelModName, modelGenName, higgsOrderMax, massList]}
{Runs the matching of the model \texttt{modelModName} onto HEFT up to Higgs
order \texttt{higgsOrderMax}. The argument \texttt{modelGenName} specifies the
corresponding \texttt{.gen} file, and \texttt{massList} contains the masses
associated with the heavy fields specified in \texttt{SetBSMFields}. If the
\texttt{.gen} file has the same name as the \texttt{.mod} file, the short form
\texttt{MatchToHEFT[modelModName, higgsOrderMax, massList]} can be used.}

\matchacommand {Options[MatchToHEFT]} {Returns the list of available options for \texttt{MatchToHEFT} which are \texttt{EFTorder -> 2} (currently the only implemented value), \texttt{SimplifyResult -> True}, \texttt{ExportDiagrams -> False}, \texttt{OutputDirectory -> Automatic}, \texttt{AlternativeBasis -> True}, and \texttt{OnlyRelevantDiagrams -> False}. These options control the EFT order, the simplification of intermediate expressions, the export of diagrams, the output directory, the export of the alternative HEFT basis, and whether only the relevant diagrams are generated.}

\matchacommand
{Results["coupling"]}
{Returns the final matching expression for the HEFT coupling specified by
\texttt{"coupling"} in the canonical HEFT basis.}

\matchacommand
{MATCHADiagrams["coupling"]}
{Returns the diagram information associated with the HEFT coupling specified by
\texttt{"coupling"}. The entries \texttt{"RelevantDiagrams"},
\texttt{"AllDiagrams"} and \texttt{"IgnoredDiagrams"} access the corresponding
diagram categories.}

\end{tcolorbox}

\captionof{table}{Summary of the main \heftmatcha commands}
\label{tab:matcha-commands}

\end{minipage}
\end{center}

\FloatBarrier

\section{Particle content of the RSE}
\label{sec:particle-content-RSE}

\begin{figure}[H]
\centering
{\scriptsize
\begin{verbatim}
M$ClassesDescription = {
S[15] == {
    SelfConjugate -> True,
    PropagatorLabel -> h,
    PropagatorType -> ScalarDash,
    PropagatorArrow -> None,
    Mass -> m,
    Indices -> {} },
V[2] == {
    SelfConjugate -> False,
    QuantumNumbers -> {Q},
    PropagatorLabel -> "W",
    PropagatorType -> Sine,
    PropagatorArrow -> Forward,
    Mass -> Mw,
    Indices -> {} },
V[3] == {
    SelfConjugate -> True,
    PropagatorLabel -> Z,
    PropagatorType -> Sine,
    PropagatorArrow -> None,
    Mass -> Mz,
    Indices -> {} },
S[3] == {
    SelfConjugate -> True,
    PropagatorLabel -> G0,
    PropagatorType -> ScalarDash,
    PropagatorArrow -> None,
    Mass -> Mz,
    Indices -> {} },
S[5] == {
    SelfConjugate -> False,
    QuantumNumbers -> {Q},
    PropagatorLabel -> GP,
    PropagatorType -> ScalarDash,
    PropagatorArrow -> Forward,
    Mass -> Mw,
    Indices -> {} },
F[1] == {
    SelfConjugate -> False,
    QuantumNumbers -> {(2*Q)/3},
    PropagatorLabel -> tq,
    PropagatorType -> Straight,
    PropagatorArrow -> Forward,
    Mass -> Mt,
    Indices -> {} },
S[6] == {
    SelfConjugate -> True,
    PropagatorLabel -> H,
    PropagatorType -> ScalarDash,
    PropagatorArrow -> None,
    Mass -> M,
    Indices -> {} }
}
\end{verbatim}
}
\caption{Selected class definitions in the FeynRules model for the RSE.}
\label{fig:rse-classes}
\end{figure}

\clearpage
\FloatBarrier
\section{Ignored diagrams for the RSE}
\label{sec:rest-diagrams}

\begin{figure}[H]
  \centering
\includegraphics[width=0.55\textwidth,
trim=0pt 100pt 0pt 0pt,
clip]{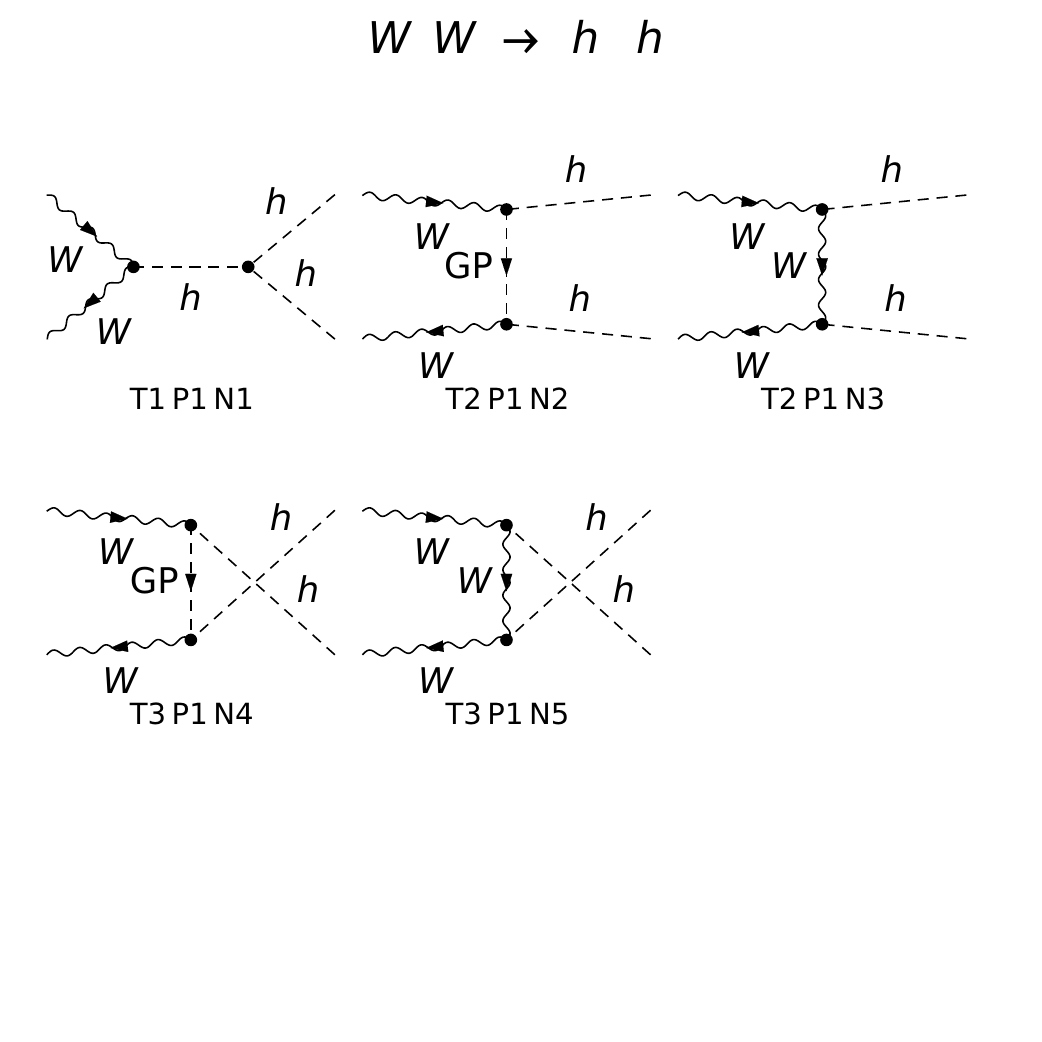}
  \caption{UV diagrams for the Real Singlet Extension involving only light particles which do not contribute to the matching of the HEFT coefficient $a_2$.}
  \label{fig:a2_ignored}
\end{figure}

\begin{figure}[H]
\centering
\includegraphics[width=0.5\textwidth]{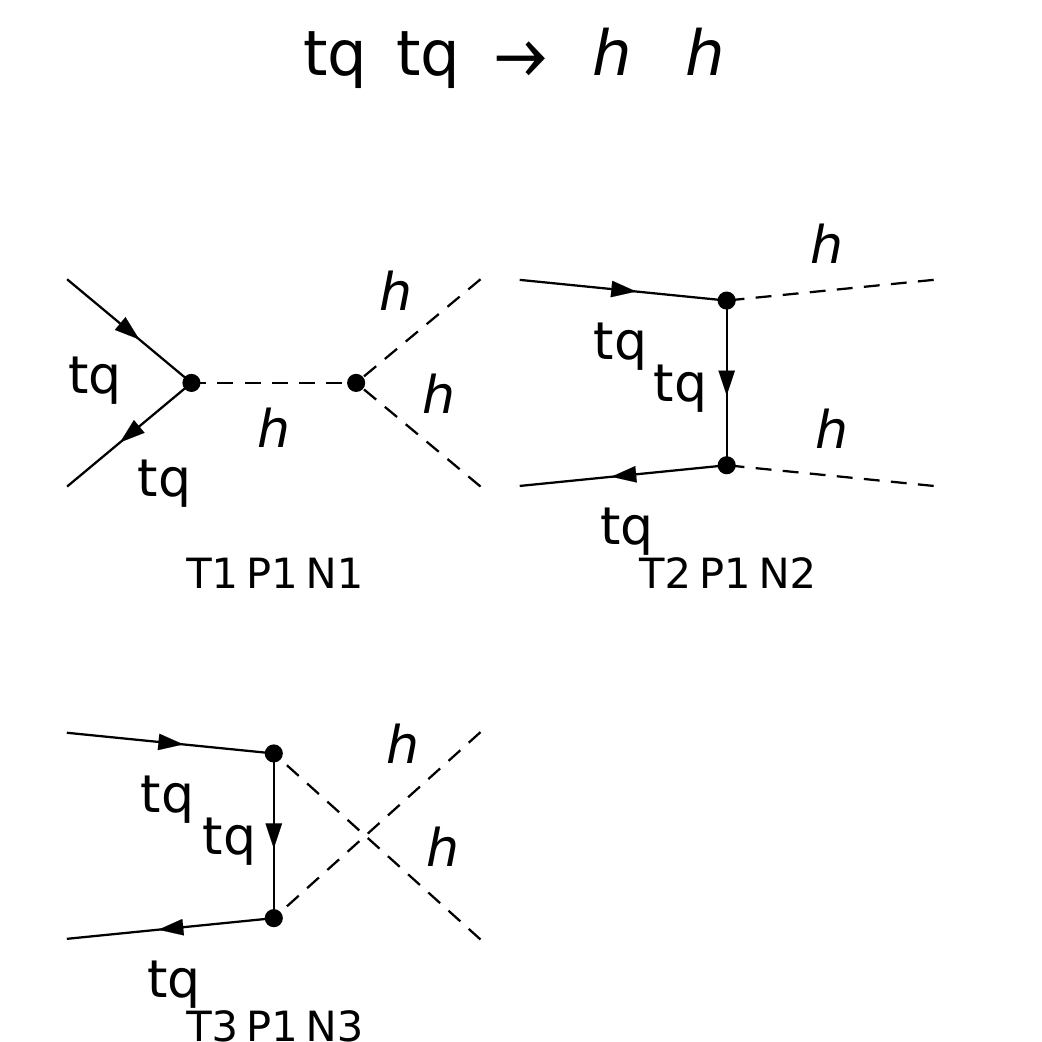}
\caption{UV diagrams for the Real Singlet Extension involving only light particles which do not contribute to the matching of the HEFT coefficient $c_2$.}
\label{fig:c2_ignored}
\end{figure}

We collect here the ignored diagrams generated by \heftmatcha for the RSE $a_2$ and $c_2$ matching channels discussed in the main text. These diagrams involve only light-particle propagators and do not contribute to the matching, as shown in Figs.~\ref{fig:a2_ignored} and~\ref{fig:c2_ignored}.

\FloatBarrier
\clearpage

\printbibliography

\end{document}